\documentclass[12pt,letterpaper]{iopart}
\usepackage{amsmath,amssymb}
\usepackage{color}
\usepackage{cite}

\usepackage[fonts]{faust}

\usepackage[linkcolor=blue, colorlinks=true, citecolor=red]{hyperref}

\newcommand\Lap[1][K]{\Delta_{\scriptscriptstyle\mathrm{#1}}}
\newcommand\dLap[1][K]{\dot\Delta_{\scriptscriptstyle\mathrm{#1}}}
\newcommand\defn[1]{{\slshape #1\/}}
\newcommand\M{\mathcal{M}}

\newcommand\U{\mathcal{U}}
\newcommand\V{\mathcal{V}}

\newcommand\X{\mathcal{X}}
\newcommand\Y{\mathcal{Y}}
\newcommand\Z{\mathcal{Z}}

\newcommand\ind{{\bullet}}

\DeclareMathOperator\Ric{Ric}
\DeclareMathOperator\Riem{Riem}

\begin{document}
\title{Perturbative stability of the approximate Killing field eigenvalue problem}
\author{Christopher Beetle and Shawn Wilder}
\address{Department of Physics, Florida Atlantic University, 777 Glades Road, Boca Raton, Florida 33431}

\begin{abstract}
An approximate Killing field may be defined on a compact, Riemannian geometry by solving an eigenvalue problem for a certain elliptic operator.  This paper studies the effect of small perturbations in the Riemannian metric on the resulting vector field.  It shows that small metric perturbations, as measured using a Sobolev-type supremum norm on the space of Riemannian geometries on a fixed manifold, yield small perturbations in the approximate Killing field, as measured using a Hilbert-type square integral norm.  It also discusses applications to the problem of computing the spin of a generic black hole in general relativity.
\end{abstract}

\section{Introduction}

The gravitational field in general relativity does not admit a local stress-energy tensor.  Physically observable quantities like energy and angular momentum therefore are defined only quasi-locally, \textit{i.e.}, as integrals over 2-dimensional surfaces in 4-dimensional spacetime.  Moreover, these quasi-local integrals are completely unambiguous only when spacetime admits a continuous symmetry generated by a Killing vector field.  Once justified in the presence of symmetry, however, the integrals can also be calculated more generally, provided that one has (a) a means of choosing a particular 2-dimensional surface over which to integrate and (b) a means of picking a particular vector field on that surface to play the role previously played by the Killing field.  The latter issue is particularly troublesome because a typical manifold has no exact symmetries whatsoever.  As a result, several schemes \cite{Dreyer, Harte, CookWhiting, Lovelace, Beetle} have appeared in the literature in recent years to define \textit{approximate} Killing vector fields on manifolds that have no real symmetries.  This paper is concerned with one such scheme, wherein the approximate Killing field arises by solving a certain eigenvalue problem on a given Riemannian geometry \cite{Matzner, Lovelace, Beetle}.  More precisely, here we investigate the \textit{stability} of this scheme under small perturbations of the geometry.  The question is important because one would expect that any physically reasonable definition of, say, the angular momentum of the gravitational field in a region of spacetime should vary continuously under a continuous deformation of that region, or of the geometry within.  We show below that the approximate Killing vector eigenfield depends continuously, in a precise sense, on the geometry used to define it.  This implies that any quasi-local integral based on this approximate Killing field varies continuously as well.

The detailed calculations that follow have an unavoidably mathematical tenor.  Nonetheless, they are motivated primarily by practical concerns arising in physical applications of the approximate Killing field eigenvalue problem.  The main such application is to the problem of calculating the spin of a black hole \cite{Lovelace,CookWhiting,Beetle}.  Let us review this application since it is the essential conceptual backdrop for the present results.

The angular momentum of a globally stationary and axisymmetric black-hole spacetime is given by the Komar formula \cite{Komar,Wald} 
\begin{equation}\label{Komar}
	J_{\mathrm{Komar}}[S, \varphi] := \frac{1}{16 \pi G} \oint_S \star \ed \varphi.
\end{equation}
The right side here is the integral of the (spacetime) Hodge dual of the exterior derivative of $\varphi_a := g_{ab}\, \varphi^b$ over a 2-sphere $S$ enclosing the black hole.  The vector field used to calculate the Komar angular momentum is a \textit{global} axial Killing field on spacetime.  For vacuum spacetimes, $J_{\mathrm{Komar}}[S, \varphi]$ is independent of the 2-sphere $S$, and coincides with the total angular momentum of spacetime measured by inertial observers at infinity.  It is natural to identify this angular momentum with the spin of the black hole in this case because the assumed symmetries preclude any gravitational radiation in the intervening space that might contribute additional angular momentum.

The integrals (\ref{Komar}) certainly exist for non-symmetric spacetimes.  But of course they depend one's choices of both $S$ and $\varphi^a$.  These ambiguities can be resolved in certain circumstances, such as if (i) one can identify the sphere $S$ with a cross-section either of the boundary of a 4-dimensional region $\mathcal{R}$ of spacetime or of a black hole's horizon, and (ii) the \textit{intrinsic} geometry on $S$ admits an axial Killing field $\xi^b$.  Various \defn{quasi-local angular momentum} formulae have been proposed for such circumstances.  Naturally, these must reduce to the Komar angular momentum (\ref{Komar}) if there exists a global axial symmetry $\varphi^a$ such that $\xi^b = \varphi^b \bigr|_S$ on the 2-sphere in question.  (The restriction $\varphi^b \bigr|_S$ must then be tangent to $S$.)  If such an extension does not exist, however, then the integrand for a quasi-local angular momentum may differ from that of (\ref{Komar}).

Consider, for example, the \defn{Brown--York angular momentum} 
\begin{equation}\label{BrownYork}
	J_{\mathrm{BY}}[S, \xi] := -\oint_{S} {}^{\scriptscriptstyle\mathcal{T}}\!\mathord{\star}\, (\xi \hook \tau).
\end{equation}
The 2-sphere $S$ here is a cross-section of a ``tubular,'' time-like, outer boundary $\mathcal{T}$ of a 4-dimensional region $\mathcal{R}$ of spacetime.  Meanwhile, ${}^{\scriptscriptstyle\mathcal{T}}\!\mathord{\star}$ denotes the 3-dimensional Hodge dual within $\mathcal{T}$, and $(\xi \hook \tau)_b := \xi^a\, \tau_{ab}$ involves the extrinsic curvature $\tau_{ab}$ of $\mathcal{T}$ within spacetime.  Another example is the \defn{generalized angular momentum} 
\begin{equation}\label{GenJ}
	J_{\mathrm{Gen}}[S, \xi] := \frac{1}{8 \pi G} \oint_S {}^{\scriptscriptstyle\Sigma}\!\mathord{\star}\, (\xi \hook K) 
\end{equation}
motivated by the study of dynamical horizons \cite{Ashtekar2003, Ashtekar2004}.  The 2-sphere $S$ in this case lies in a given partial Cauchy slice $\Sigma$, and is such that the expansion of one of the two null geodesic congruences emanating from it vanishes.  Meanwhile, ${}^{\scriptscriptstyle\Sigma}\!\mathord{\star}$ denotes the 3-dimensional Hodge dual within $\Sigma$, and $K_{ab}$ denotes the extrinsic curvature of $\Sigma$ as it is embedded in spacetime.

The common feature of (\ref{BrownYork}) and (\ref{GenJ}) is that the sphere $S$ is fixed geometrically by features of the surrounding spacetime.  Both formulae are then on their surest footing when the intrinsic geometry on $S$ admits an axial Killing field $\xi^a$.  To continue interpreting (\ref{BrownYork}) or (\ref{GenJ}) as \textit{the} angular momentum of a black hole when $S$ is not axially symmetric, one must specify how to construct a similarly geometrically preferred vector field $\xi^a$ from an arbitrary 2-sphere metric.  Moreover, this vector field must reduce to the axial Killing field whenever one exists.  A natural way to do this is to seek a best approximation, in some prescribed sense, to a Killing field on $S$.  Several schemes of exactly this type have been proposed \cite{Matzner, CookWhiting, Lovelace, Beetle, Dreyer, Harte}.

In principle, prescribing the two structures (a) and (b) gives a precise analytic formula for a black hole's angular momentum.  But subtleties can arise in practical applications, such as in numerical relativity.  In particular, the preferred 2-sphere $S$ can be expensive to calculate numerically \cite{Thornburg}.  Thus, the sphere $S$ one actually uses in practice may be displaced slightly, and have a slightly different intrinsic geometry, from the ideal sphere on which the analytic formula is based.  If, in addition, the choice of approximate Killing field depends on some \textit{ad hoc} choices, such as in \cite{Dreyer,Harte}, then one must also worry about how those choices  affect the angular momentum.  Although the approximate Killing field derived from the eigenvalue problem described below does not suffer from the latter ambiguity, it is not immediately clear how important it is to locate the desired $S$ exactly in spacetime.

This paper shows explicitly that small errors in the ideal intrinsic geometry on $S$ produce only small errors in the approximate Killing field determined by the eigenvalue problem described below.  Importantly, the quasi-local angular momentum formulae (\ref{BrownYork}) and (\ref{GenJ}) both lead to expressions like  
\begin{equation}
	J[S] = \oint_S \xi^a[g]\, j^b\, g_{ab}\, \epsilon.
\end{equation}
This is an $L^2$ inner product of an approximate Killing field $\xi^a[g]$, constructed (non-locally) from the intrinsic geometry $g_{ab}$ of $S$, with a vector field $j^b$ that depends (locally) on both the intrinsic and extrinsic geometry of $S$.  The change of any such integral under perturbations of the fields involved can be bounded using the Schwarz inequality.  The question thus arises whether the $L^2$ norm of the perturbation in $\xi^a[g]$ can be bounded in terms of perturbations of the intrinsic metric itself, and therefore in terms of potential errors locating $S$ exactly within spacetime.  We show below that it can.

Although our primary motivation is the application to the 2-dimensional geometry of the horizon of a black hole, the results below actually apply to perturbations of a Riemannian geometry of arbitrary dimension.  The topology of the manifold, however, must remain compact and without boundary.  One could presumably relax the latter restriction by imposing appropriate boundary conditions, but we will not explore this question in detail.

The outline of the paper is as follows.  Section 2 reviews the eigenvalue problem from \cite{Matzner, Lovelace, Beetle} whose solution yields the approximate Killing field.  Section 3 describes a novel scheme to fix the (global) normalization of that vector eigenfield.  Section 4 discusses perturbations of the eigenvalue problem, and shows that perturbations in the eigenfields can be bounded if the perturbation of the underlying operator can be bounded appropriately.  Section 5 shows that the perturbed operator can indeed be bounded in this way.  We conclude in Section 6 with some comments and discussion.

\section{The Killing Laplacian and approximate Killing fields}

Let $\mathcal{M}$ be a compact Riemannian manifold with metric $g_{ab}$, and let $u^a$ denote an arbitrary vector field on that manifold.  Define the action integral 
\begin{equation}\label{killact}
	\mathcal{S}[u](\kappa) 
		:= \frac{1}{2} \int_{\mathcal{M}} \Lie_u g_{ab}\, \Lie_u g_{cd}\, g^{ac}\, g^{bd}\, \epsilon
			+ \kappa \biggl( 1 - \int_{\mathcal{M}} u^a\, u^b\, g_{ab}\, \epsilon \biggr),
\end{equation}
where $\epsilon$ denotes the volume form on $\mathcal{M}$ derived from $g_{ab}$.  Minimizing this integral obviously minimizes the Lie derivative of the metric in an averaged sense over the whole of $\mathcal{M}$, subject to a constraint that the natural $L^2$ norm of the vector field $u^a$ on $\mathcal{M}$ remains fixed.  This rules out the limit $u^a \to 0$, where clearly  (\ref{killact}) always has a trivial minimum.   The Euler--Lagrange equation for (\ref{killact}) is 
\begin{equation}\label{killlap}
	\Lap\, u^d := - 2\, \delta^{(d}_c\, g^{a)b}\, \grad\!_a\, \grad\!_b\, u^c = \kappa_u\, u^d.
\end{equation}
Its solutions are vector eigenfields $u^a$ for the \defn{Killing Laplacian} operator $\Lap$ with corresponding eigenvalues $\kappa_u$.  Historically, it appears that this operator was first introduced to the relativity literature by Matzner \cite{Matzner}, and independently to the differential geometry literature by Bochner and Yano \cite{Yano}.  More recently, Owen and collaborators \cite{Lovelace} have reintroduced the idea in specific application to the angular momentum of black holes.

The Killing Laplacian $\Lap$ has several important characteristics that dictate the nature of its eigenvalue problem.  First, it is positive and Hermitian on the Hilbert space $L^2(\M^\ind, g)$ of vector fields\relax
\footnote{The superscript on $\M$ indicates the tensorial structure of the fields in the Hilbert space $L^2(\M^\ind, g)$, and we include $g$ explicitly to emphasize that the inner product varies with the metric.  We will drop the latter notation shortly, but keep the former throughout because several Hilbert spaces of different types of tensor fields appear in our calculations.}
on $\M$ that are square integrable with respect to the natural measure defined by $g_{ab}$.  The eigenvalues $\kappa_u$ are therefore real and positive, and the corresponding eigenfields $u^a$ are real and orthogonal in the inner product 
\begin{equation}\label{ipL2v}
	\iprod{u}{v}_{L^2(\M^\ind, g)} := \oint_\M \bar u^a\, v^b\, g_{ab}\, \epsilon.
\end{equation}
Second, the Killing Laplacian is (strongly) elliptic \cite{McLean}.  Such operators have a \textit{complete} basis of \textit{normalizable} eigenfields $u^a$ with \textit{discrete} eigenvalues $\kappa_u$ that are \textit{unbounded} above.

We define the \defn{approximate Killing vector field} for an arbitrary Riemannian geometry $(\M, g_{ab})$ to be simply the eigenfield $u_0^a$ of the Killing Laplacian with the smallest eigenvalue $\kappa_0 \ge 0$.  If $g_{ab}$ admits an actual Killing field $\xi^a$, then $u_0^a$ coincides with it and has eigenvalue $\kappa_0 = 0$.  But of course the definition is valid generally.  It also may happen that the lowest eigenvalue $\kappa_0$ of $\Lap$ is degenerate, and $u_0^a$ therefore may not be unique.  (This happens, for example, for a round sphere, in which case the lowest eigenvalue is $\kappa_0 = 0$ and the corresponding eigenspace is 3-dimensional.  We do not have a similarly concrete example of a geometry with $\kappa_0 > 0$ for which the corresponding eigenspace is multi-dimensional, but there is no obvious reason to suppose that such geometries do not exist.)  We will not consider this possibility explicitly below, however, because the generic results for non-degenerate $\kappa_0$ that we do present generalize straightforwardly to this exceptional case.

\section{Normalizing the approximate Killing field on a sphere}

One of course has to normalize the approximate Killing field $u_0^a$ appropriately in order to calculate a definite spin for a black hole.  As for any eigenfield, the only freedom is to scale it by a constant over all of $\M$.  But it is not immediately clear how to choose that constant scaling.  The normalization for which the constraint term vanishes in the action principle (\ref{killact}) just fixes the Hilbert norm of $u^a$ in $L^2(\M^\ind, g)$.  This is almost certainly not the correct normalization if we are trying to model an axial Killing field on a topological 2-sphere.  (It would be a fine normalization for a translational Killing field on a torus, say, but this is not the case that interests us most.)  The question therefore arises how to normalize $u_0^a$ when $\M \sim S^2$ such that it reduces to the standard normalization for a axial Killing field when the sphere's geometry happens to be axially symmetric.

Owen and collaborators \cite{Lovelace} outline such a normalization scheme for the case when $u_0^a$ is constrained to be divergence free on $\M \sim S^2$, and is assumed to vanish only at two isolated points.  The orbits of the flow $U_0(t)$ generated by $u_0^a$ are then all circles, each of which closes for some, though not necessarily the same, value of $t$.  The approximate Killing field is normalized such that a certain average of those parameter values $t$ at which each orbit closes is $2 \pi$.  This reduces to the correct normalization when $u_0^a = \xi^a$ is a genuine axial Killing field, whose orbits all close at $t = 2 \pi$ by definition.

Here we propose an alternate normalization for the approximate Killing field on a topological 2-sphere.  Our proposal is (a) more local in that it does not require a global average over an entire sphere, (b) does not need to restrict to divergence-free vector fields, and (c) does not need to assume that $u_0^a$ has exactly two isolated zeroes.

Suppose for the moment that $(\M \sim S^2, g_{ab})$ admits an axial Killing field $\xi^a$.  The defining characteristic of an \textit{axial} Killing field is that it vanishes at at least one $p_0 \in \M$.  The flow $\Xi(t)$ generated by $\xi^a$ has period $t = 2 \pi$ near any such $p_0$ if and only if 
\begin{equation}\label{dxinorm}
	\bigl[ \grad\!_a\, \xi^b \cdot \grad\!_b\, \xi^a \bigr]_{p_0} = -2.
\end{equation}
This is so because the flow near $p_0$ generates a corresponding flow in the tangent space $T_{p_0} \M$, which is a family of SO(2) rotations for the \textit{Euclidean} metric $g_{ab}(p_0)$ thereon.  This family of SO(2) rotations closes at $t = 2 \pi$ if and only if its matrix generator is properly normalized, which is precisely the condition (\ref{dxinorm}).  We show in a companion paper \cite{axial} that this convention to normalize $\xi^a$ at \textit{one} fixed point $p_0$ necessarily selects the desired normalization of the Killing field throughout.  Moreover, this result is independent of which $p_0$ one uses.

It is well known that any vector field, and in particular the approximate Killing field $u_0^a$ for an arbitrary 2-sphere metric $g_{ab}$, must vanish at at least one $p_0 \in \M$.  We therefore can choose the constant scaling of $u_0^a$ throughout $\M$ such that 
\begin{equation}\label{du0norm}
	\bigl[ \grad\!_a\, u_0^b \cdot \grad^a\, u_{0b} \bigr]_{p_0} = 2 
\end{equation}
at that one point.  If there is more than one such point, then one may either just pick one, though then the normalization is unlikely to be independent of that choice, or else average the left side of (\ref{du0norm}) over all such points.  (It may be necessary to assume that there are only finitely many zeroes $p_0 \in \M$ of $u_0^a$, which certainly seems likely for the lowest eigenfields of an elliptic operator.  We have no proof of this, however.)  Whichever approach one chooses, the normalization of $u_0^a$ given by (\ref{du0norm}) has both of the desired properties that (a) it exists for any Riemannian 2-sphere geometry, and (b) it gives the correct scaling of $u_0^a = \xi^a$ when that geometry happens to be axially symmetric.

\section{Perturbations of the Killing Laplacian and its eigenfields}

Next we want to show that the approximate Killing vector eigenvalue problem is \textit{stable}, that is, that small changes in $g_{ab}$ induce only small changes in $u_0^a$.  We approach this problem technically by considering a smooth, one-parameter family $\bigl( \M, g_{ab}(\lambda) \bigr)$ of Riemannian geometries on a fixed, compact manifold $\M$ without boundary.

As $g_{ab}(\lambda)$ varies with $\lambda$, so will any geometric structure derived from it.  The relevant formulae for perturbations of these derived structures are the standard ones from linear perturbation theory in general relativity (see, \textit{e.g.}, \cite{Wald}).  To summarize our conventions, the Levi-Civita connection $\grad\!_a$ changes according to 
\begin{equation}\label{dConn}
	\delta \bigl( \grad\!_a\, \omega_b \bigr) 
		= \dot\grad\!_{ab}{}^c\, \omega_c 
		:= \frac{1}{2}\, g^{cd}\, \bigl( \grad\!_d\, \dot g_{ab} - 2\, \grad\!_{(a}\, \dot g_{b)d} \bigr)\, \omega_c, 
\end{equation}
where $\omega_b$ is an arbitrary fixed (\textit{i.e.}, $\lambda$-independent) covector field, while the inverse metric $g^{ab}$ and the volume element $\epsilon$ change according to 
\begin{equation}\label{dinvvol}
	\delta \bigl( g^{ab} \bigr) = - \dot g^{ab} := - g^{ac}\, g^{bd}\, \dot g_{cd}
	\qquad\text{and}\qquad
	\delta\epsilon = \frac{1}{2}\, \dot g\, \epsilon := \frac{1}{2}\, g^{ab}\, \dot g_{ab}\, \epsilon, 
\end{equation}
respectively.  We denote perturbations using either the operator $\delta := \pdby{\lambda} \bigr|_{\lambda = 0}$, when necessary, or equivalently using a dot accent.  Using these standard results, the perturbation of the Killing Laplacian (\ref{killlap}) takes the form 
\begin{equation}\label{dkilllap}
	\dLap v^a 
		= \underbrace{\bigl( 2\, \delta^{(a}_d\, \dot g^{b)c} \bigr)}_{X^{abc}{}_d} \grad\!_b \grad\!_c\, v^d
			+ \underbrace{( 2\, \dot\grad^b{}_c{}^a 
				+ g^{ab}\, \dot\grad\!_{cd}{}^d
				- \delta_c^a\, \dot\grad\!_d{}^{db} )}_{Y^{ab}{}_c} \grad\!_b\, v^c
			+ \underbrace{\bigl( 2\, \grad\!_c \dot\grad\!_b{}^{(ac)} \bigr)}_{Z^a{}_b{}} v^b, 
\end{equation}
where $v^d$ is a fixed vector field.  If the vector field is not fixed, such as if it is an eigenfield $u_n^a(\lambda)$ of the Killing Laplacian $\Lap(\lambda)$ constructed from the metric $g_{ab}(\lambda)$, then of course there are additional terms coming from the variation of the vector field itself.

The process of relating the perturbation (\ref{dkilllap}) of the Killing Laplacian to the perturbations of its eigenfields is more or less the familiar one from elementary quantum mechanics.  Perturbing the eigenvalue problem (\ref{killlap}) gives 
\begin{equation}\label{deveq}
	\bigl( \kappa_n - \Lap \bigr) \dot u_n^a = \bigl( \dLap - \dot\kappa_n \bigr) u_n^a, 
\end{equation}
which formally determines $u_n^a$ up to its component in the eigenspace of $\Lap$ containing $u_n^a$ itself.  One difference from the usual quantum mechanical perturbation theory is that the inner product on the Hilbert space $L^2(\M^\ind, g)$ varies with $\lambda$, too.  Imposing the usual normalization condition leads to 
\begin{equation}
	\norm{u_n(\lambda)}^2_{L^2(\M^\ind, g(\lambda))} = 1
	\qquad\leadsto\qquad
	2 \real \iprod{u_n}{\dot u_n}_{L^2(\M^\ind, g)} = - \iprod{u_n}{\mathcal{A} u_n}_{L^2(\M^\ind, g)}, 
\end{equation}
where $\mathcal{A}$ denotes the algebraic operator on $L^2(\M^\ind, g)$ derived from the tensor field 
\begin{equation}\label{alphadot}
	\dot\alpha^a{}_b := \dot g^a{}_b + \frac{1}{2}\, \dot g\, \delta^a_b.
\end{equation}
This accounts for the changes of the point-wise inner product in the tangent space and the volume element in the Hilbert-space inner product (\ref{ipL2v}).  Collecting these results, and keeping $u_n^a$ real-valued, gives the formal expression 
\begin{equation}\label{dunform}
	\dot u_n^a = \biggl( P_n^\shortperp\, \frac{1}{\kappa_n - \Lap}\, P_n^\shortperp \biggr)\, \dLap\, u_n^a 
		- \frac{1}{2}\, \dot\alpha^a{}_b\, u_n^b
\end{equation}
for the perturbation of the $n^{\mathrm{th}}$ eigenfield of the Killing Laplacian.

The operator in braces in (\ref{dunform}) is the pseudoinverse of the operator on the left side of (\ref{deveq}).  That is, it represents convolution with the Green function for the restriction of that operator to the subspace of $L^2(\M^\ind, g)$ orthogonal to the eigenspace of $\Lap$ with eigenvalue $\kappa_n$.  We use $P_n^\shortperp$ to denote the projection onto that orthogonal subspace.  The pseudoinverse operator is bounded (in fact, compact), and its operator norm obeys 
\begin{equation}\label{pinorm}
	\norm{P_n^\shortperp\, \frac{1}{\kappa_n - \Lap}\, P_n^\shortperp} \le \frac{1}{\gamma_n}, 
	\qquad\text{where}\qquad
	\gamma_n := \min (\kappa_n - \kappa_{n-1}, \kappa_{n+1} - \kappa_n)
\end{equation}
is the gap around $\kappa_n$ in the spectrum of $\Lap$.  (In the case $n = 0$ of primary interest, its definition is simply $\gamma_0 := \kappa_1 - \kappa_0$.)  We are relying on general results for elliptic operators on compact manifolds \cite{McLean,ChoquetBruhat} to show that that spectrum is discrete, and therefore that this bound on the operator norm of the pseudoinverse is meaningful.  The important point, however, is that bounding the perturbation $\dot u_n^a$ of the eigenfield in Hilbert space amounts to bounding the action of the perturbed Killing Laplacian $\dLap$ on the unperturbed $u_n^a$.

\section{Bounding the perturbed eigenfields of the Killing Laplacian}

This section shows that the action $\dLap v^a$ of the perturbed Killing Laplacian on an arbitrary vector field is necessarily bounded as long as both $\Lap v^a$ and $v^a$ itself have finite $L^2$ norm.  (This will be the case, for example, if $v^a$ lies in the domain of $\Lap$ in Hilbert space.)  This implies further that the norm of the perturbation $\dot u_n^a$ of any eigenfield of the Killing Laplacian is less than a certain constant $C_n$ times the norm of the unperturbed $u_n^a$.  Moreover, this $C_n$ scales in direct proportion to the metric perturbation $\dot g_{ab}$.  For any given $n$, we therefore can make (the norm of) $\dot u_n^a$ as small as we like by taking $\dot g_{ab}$ sufficiently small.  In other words, each eigenfield of the Killing Laplacian depends \textit{continuously}, in the natural $L^2$ topology on Hilbert space, on the Riemannian metric $g_{ab}$.  This gives the precise mathematical sense in which the approximate Killing field determined by the eigenvalue problem (\ref{killlap}) is stable.

The following proof of these claims is organized in three steps.  The first is just to use the triangle inequality and certain point-wise bounds on the tensor fields $X^{ab}{}_c{}^d$, $Y^b{}_c{}^d$ and $Z_c{}^d$ from (\ref{dkilllap}) to show that 
\begin{equation}\label{dLapTri}
	\norm[\big]{\dLap v}_{L^2(\M^\ind)} \le 
		C_X\, \norm{\grad\grad v}_{L^2(\M_{\ind\ind}{}^\ind)}
		+ C_Y\, \norm{\grad v}_{L^2(\M_\ind{}^\ind)}
		+ C_Z\, \norm{v}_{L^2(\M^\ind)}.
\end{equation}
The new $L^2$ norms appearing here are just the natural ones for tensor fields with the given index structure under the \textit{background} metric $g_{ab}(0)$, as in (\ref{ipL2v}).  (All norms and inner products from this point forward derive from the background metric, which accordingly we no longer denote explicitly.)  The constants $C_X$, $C_Y$ and $C_Z$ depend on the tensor fields from (\ref{dkilllap}), and scale in direct proportion to the metric perturbation $\dot g_{ab}$.  We derive (\ref{dLapTri}) in subsection \ref{dLap<Hess}, including specific values for $C_X$, $C_Y$ and $C_Z$.

The most divergent term on the right side of (\ref{dLapTri}) typically will be the first, which involves the (covariant) Hessian $\grad\!_a \grad\!_b\, v^c$ of the vector field.  The second step of our calculation is therefore to bound this Hessian such that 
\begin{align}\label{HessKill}
	\norm{\grad\grad v}^2_{L^2(\M_{\ind\ind}{}^\ind)} \le 
		\norm{\Lap v}^2_{L^2(\M^\ind)}
		&+ \norm{\Riem v}^2_{L^2(\M_{\ind\ind}{}^\ind)}
		\notag\\&\hspace{2em}
		+ C_U\, \norm{\grad v}^2_{L^2(\M_\ind{}^\ind)} 
		+ C_V\, \norm{v}^2_{L^2(\M^\ind)}.
\end{align}
The second term on the right here denotes the norm of the tensor field $v^a\, R_{abc}{}^d$, which actually is equal to the Hessian $\grad\!_b \grad\!_c\, \xi^d$ if $v^a = \xi^a$ happens to be a Killing field.  This provides a nice check on this inequality since the first term on the right vanishes in that case, and one has only some very general bounds on the constants $C_U$ and $C_V$ in the second line.  Indeed, those constants arise in much the same way as the corresponding constants in (\ref{dLapTri}), though they depend only on the background geometry $g_{ab}$ and not its perturbation $\dot g_{ab}$.  We derive (\ref{HessKill}) in subsection \ref{Hess<Lap}.

The first two steps of the calculation show that the action of the perturbed Killing Laplacian $\dLap$ on an arbitrary vector field $v^a$ has bounded $L^2$ norm if we can control (a) the $L^2$ norm of $\Lap v^a$, and (b) the $L^2$ norm of the first-order derivatives $\grad\!_a\, v^b$.  The third step of the calculation shows that the first-order derivatives can also be bounded in terms of $\Lap v^a$ according to 
\begin{equation}\label{GradLap}
	\norm{\grad v}^2_{L^2(\M_\ind{}^\ind, g)} \le 
		\iprod{v}{\Lap v}_{L^2(\M^\ind, g)} 
		+ C_{\Ric}\, \norm{v}^2_{L^2(\M^\ind, g)}, 
\end{equation}
where $C_{\Ric}$ bounds the linear operator $v^a \mapsto R^a{}_b\, v^b$ on Hilbert space derived from the Ricci tensor of the background metric.  Again, it is independent of the perturbation $\dot g_{ab}$.  We establish this last result in subsection \ref{Grad<Lap}.

At this stage we (will) have shown that, although the perturbed Killing Laplacian $\dLap$ is certainly an unbounded operator on Hilbert space, the ``unbounded part'' of its action on any given vector field $v^a$ is \textit{relatively} bounded by the action $\Lap v^a$ of the Killing Laplacian itself.  (One could insert some technical language at this point regarding the domains of these operators, but in fact this argument shows that it is natural simply to choose the domain of $\dLap$ to coincide with that of $\Lap$.)  The action $\Lap u_n^a = \kappa_n\, u_n^a$ of the Killing Laplacian on any of its own eigenfields is certainly bounded.  We therefore conclude from (\ref{dunform}) that the perturbation $\dot u_n^a$ of any eigenfield of $\Lap$ satisfies
\begin{equation}\label{unBound}
	\norm{\dot u_n}_{L^2(\M^\ind)} \le C_n\, \norm{u_n}_{L^2(\M^\ind)}
\end{equation}
for a constant $C_n$ depending on the unperturbed eigenvalue $\kappa_n$, the unperturbed spectral gap $\gamma_n$ around that eigenvalue, and a number of constants $C$ like those described above depending on $g_{ab}$ and $\dot g_{ab}$, but not on $n$.  We establish this result in detail in subsection \ref{dLap<Lap}, where we give a specific formula for the $C_n$.

\subsection{The Hessian and lower-order terms bound the perturbed Killing Laplacian}
\label{dLap<Hess}

Applying the triangle inequality to the perturbed Killing Laplacian (\ref{dkilllap}) gives 
\begin{align}\label{dLtri}
	\norm[\big]{\dLap v}_{L^2(\M^\ind)} 
		:={}&{} \norm{\X \grad \grad v + \Y \grad v + \Z v}_{L^2(\M^\ind)}
			\notag\\[1ex]
		\le{}&{} \norm{\X \grad \grad v}_{L^2(\M^\ind)} 
			+ \norm{\Y \grad v}_{L^2(\M^\ind)} 
			+ \norm{\Z v}_{L^2(\M^\ind)}, 
\end{align}
where $\X$, $\Y$ and $\Z$ denote the obvious algebraic operators on Hilbert space derived from the tensor fields $X$, $Y$ and $Z$ in (\ref{dkilllap}), respectively.

The next step is more or less the same for each of the three terms in (\ref{dLtri}).  Let's consider the first in detail, which is 
\begin{equation}\label{dLXadj}
	\norm{\X \grad \grad v}^2_{L^2(\M^\ind)}
		= \iprod{\grad \grad v}{\X^\dagger \X \grad \grad v}_{L^2(\M_{\ind \ind}{}^\ind)}.
\end{equation}
The operator $\X^\dagger \X$ is obviously non-negative.  It is also bounded because $\M$ is compact by assumption and both $\X$ and $\X^\dagger$ act algebraically.  To see this, note that the tensor field underlying $\X^\dagger$ is just the point-wise adjoint of that underlying $\X$ itself: 
\begin{equation}
	X^\dagger{}_{ab}{}^c{}_d\, w^d := g_{aj}\, g_{bk}\, g^{cl}\, X^{ijk}{}_l\, g_{id}\, w^d
	\qquad\text{or}\qquad
	X^\dagger{}^A{}_d := g^{AB}\, g_{cd}\, X^c{}_B.
\end{equation}
We use uppercase  indices in the latter expression to denote ``vectors'' in the appropriate tensor product of tangent and co-tangent spaces at each point of $\M$.  The product $X^\dagger{}^A{}_b\, X^b{}_C$ at each $p \in \M$ is then a non-negative, Hermitian map from that tensor product space to itself.  In fact, each of these maps must have a substantial kernel because $X^i{}_J$ itself maps each eigenvector of $(X^\dagger X){}^A{}_C$ to an eigenvector of $(X X^\dagger){}^a{}_c$ having the same eigenvalue, and conversely $X^\dagger{}^I{}_j$ maps eigenvectors in the opposite direction.  It follows that each $(X^\dagger X){}^A{}_C$ has at most as many non-zero eigenvalues as the dimension of $\M$, and each of these is also an eigenvalue of $(X X^\dagger){}^a{}_c$ at the same point.  Define $M(p)$ to be the largest such eigenvalue at each $p \in \M$.  This function must be smooth since the tensor fields used to define it are smooth.  It therefore attains a finite maximum value on the compact manifold $\M$.  This ``maximum maximum'' eigenvalue bounds the Hilbert space operator $\X^\dagger \X$ such that 
\begin{equation}\label{Xbound}
	\norm{\X \grad \grad v}_{L^2(\M^\ind)} 
		\le C_X\, \norm{\grad \grad v}_{L^2(\M_{\ind\ind}{}^\ind)}.
\end{equation}
We have shown that the bounding constant $C_X$ here is given by  
\begin{align}\label{CXdef}
	C_X^2 := \max_{p \in \M} M(p)
		:={}&{} \max_{p \in \M} \max_{s \in S_p \M} \Bigl[ s_a\, X^{abc}{}_d\, X_{ebc}{}^d\, s^e \Bigr](p)
			\notag\\[1ex]
		\le{}&{} \max_{p \in \M} \Bigl[ X^{abc}{}_d\, X_{abc}{}^d \Bigr](p), 
\end{align}
where we have used the metric to raise and lower indices and the inner maximum on the first line is over the unit sphere $S_p \M$ in each tangent space $T_p \M$.  The final inequality arises because the largest eigenvalue at each point is less than the sum of all the (non-negative) eigenvalues, which of course is just the trace of the operator.  This trace might be simpler to calculate and maximize over $\M$ in practice.

We can bound the other two terms on the right side of (\ref{dLtri}) in a completely analogous manner.  The overall bound (\ref{dLapTri}) then follows directly.  Note that the tensor fields $X^{abc}{}_d$, $Y^{ab}{}_c$ and $Z^a{}_b$ from (\ref{dkilllap}) are all linear in the perturbation $\dot g_{ab}$ of the metric.  It follows that the bounding constants $C_X$, $C_Y$ and $C_Z$ in (\ref{dLapTri}) all scale homogeneously under a reparameterization of the family of geometries $g_{ab}(\lambda)$ underlying the perturbation.  The left side of (\ref{dLapTri}) scales in the same manner, of course, so the result holds regardless of how the parameter $\lambda$ is defined geometrically.

\subsection{The Killing Laplacian and lower-order terms bound the Hessian}
\label{Hess<Lap}

We begin by establishing a couple local differential identities involving the Killing Laplacian and related operators.  The first is a Weitzenb\"ock identity relating $\Lap$ to the \defn{Laplace--Beltrami operator} $\Lap[B] := - \grad\!_a \grad^a$ acting on vector fields: 
\begin{align}\label{KLWeitz}
	\Lap v^a 
		:= - \grad\!_b \grad^b\, v^a - \grad\!_b \grad^a\, v^b
		&= \Lap[B] v^a - \grad^a \grad\!_b\, v^b - R^a{}_b\, v^b
			\notag\\[1ex]
		&= (\Lap[B] v - \Grad \Div v - \Ric v)^a.
\end{align}
The index-free notation on the second line will be useful for the manipulations in Hilbert space that follow.  The second set of results we will need is 
\begin{equation}\label{ddcvf}
	0 	= 2\, \grad\!_a \grad\!_b \grad^{[a} v^{b]} 
		= \Div (\ed^\dagger \ed v)
		= \Div (\Lap[B] v + \Grad \Div v + \Ric v), 
\end{equation}
where we use the metric to raise and lower indices as needed, $\ed$ in the third expression denotes the exterior derivative, and $\ed^\dagger$ denotes its adjoint in the natural Hilbert inner product on the space of 2-form fields.  Note that the final equality also holds without the initial divergence, as a relation between vector fields.

We now compute the $L^2$ norm of $\Lap v^a$ as follows.  First, substitute for $\Lap$ using (\ref{KLWeitz}) and expand.  Second, use integrations by parts and the identities (\ref{ddcvf}) to simplify the resulting cross-terms, leaving 
\begin{align}\label{KBLnorm}
	\norm{\Lap v}^2_{L^2(\M^\ind, g)}
		= \norm{\Lap[B] v}^2_{L^2(\M^\ind)} 
			&- 2 \real \iprod{\ed v}{\ed \Ric v}_{L^2(\M_{[\ind\ind]})}
			\notag\\&\hspace{2em}
			+ 3\, \norm{\Grad \Div v + \Ric v}^2_{L^2(\M^\ind)}.
\end{align}
Note that the second term on the right here involves at most first-order derivatives of $v^a$.  It can also be written in the form 
\begin{align}\label{KBLcross}
	\iprod{\ed v}{\ed \Ric v}_{L^2(\M_{[\ind\ind]})}
		:={}&{} \frac{1}{2} \oint_\M \bigl( 2\, \grad^{[a}\, \bar v^{b]} \bigr) 
			\bigl( 2\, \grad\!_{[a} R_{b]c} \cdot v^c + 2\, R_{d[b}\, g_{a] c}\, \grad^c\, v^d \bigr)\, \epsilon
			\\[1ex]\notag
		={}&{} \oint_\M \bigl( 2\, g^{c [a}\, R^{b] d} \bigr) 
			\bigl( \grad\!_a\, \bar v_b \bigr) \bigl( \grad\!_c\, v_d \bigr)\, \epsilon
		- \oint_\M \bigl( \grad\!_d\, R_{abc}{}^d \bigr) \bigl( \grad^a\, \bar v^b \bigr)\, v^c\, \epsilon, 
\end{align}
where we have used the partially contracted Bianchi identity $2\, \grad\!_{[a}\, R_{b]c} = - \grad\!_d\, R_{abc}{}^d$
in the second line.  Third, the squared norm of $\Lap[B] v^a$ is related to the squared norm of the Hessian of $v^a$ by additional lower-order derivative terms: 
\begin{align}\label{LBLHess}
	\norm{\Lap[B] v}^2_{L^2(\M^\ind)}
		&= \norm{\grad \grad v}^2_{L^2(\M_{\ind\ind}{}^\ind)}
			+ \oint_\M \bigl( R^{ac}\, g^{bd} \bigr) 
				\bigl( \grad\!_a\, \bar v_b \bigr) \bigl( \grad\!_c\, v_d \bigr)\, \epsilon
			\notag\\&\hspace{2em}
			+ \oint_\M \bigl( \grad\!_d\, R_{bca}{}^d \bigr) \bigl( \grad^a\, \bar v^b \bigr)\, v^c\, \epsilon
			- \norm[\big]{\Riem v}^2_{L^2(\M_{\ind\ind}{}^\ind)}.
\end{align}
The last term here is the squared $L^2$ norm of $v^a\, R_{abc}{}^d$.  As mentioned above, this is equal to the Hessian $\grad\!_b \grad\!_c\, \xi^d$ when $v^a = \xi^a$ is a Killing field.  Finally, insert (\ref{LBLHess}) and (\ref{KBLcross}) into (\ref{KBLnorm}) and collect terms to find 
\begin{align}\label{bndHess}
	\norm{\grad \grad v}^2_{L^2(\M_{\ind\ind}{}^\ind)}
		&\le \norm{\grad \grad v}^2_{L^2(\M_{\ind\ind}{}^\ind)} 
			+ 3\, \norm{\Grad \Div v + \Ric v}^2_{L^2(\M^\ind)}
			\notag\\[1ex]
		&= \norm{\Lap v}^2_{L^2(\M^\ind)}
			+ \norm{\Riem v}^2_{L^2(\M_{\ind\ind}{}^\ind)}
			- \oint_\M \underbrace{\bigl( \grad\!_a \grad\!_c\, R^{abcd} \bigr)}_{V^{bd}}
				\bar v_b\, v_d\, \epsilon
			\notag\\[-2ex]&\hspace{2em}
			+ \oint_\M \overbrace{\bigl( 2\, g^{ac}\, R^{bd} - R^{ac}\, g^{bd} 
					- g^{ad}\, R^{cb} - g^{cb}\, R^{ad} \bigr)}^{U^{abcd}}
				\bigl( \grad\!_a\, \bar v_b \bigr) \bigl( \grad\!_c\, v_d \bigr)\, \epsilon
			\notag\\[1ex]
		&\le \norm{\Lap v}^2_{L^2(\M^\ind)} 
			+ \norm{\Riem v}^2_{L^2(\M_{\ind\ind}{}^\ind)}
			\notag\\&\hspace{2em}
			+ \norm{\grad v}_{L^2(\M_\ind{}^\ind)}\, 
				\norm{\U \grad v}_{L^2(\M_\ind{}^\ind)}
			+ \norm{v}_{L^2(\M^\ind, g)}\, \norm{\V v}_{L^2(\M^\ind)}.
\end{align}
The last line of this calculation uses the Schwarz inequality.  The operators $\U$ and $\V$ on Hilbert space are defined in the obvious way from the tensor fields $U_a{}^{bc}{}_d$ and $V^b{}_d$, respectively.  These algebraic operators are bounded, and their bounds can be computed by the same argument that led to (\ref{Xbound}).  Our main result (\ref{HessKill}) follows.  Unlike (\ref{dLapTri}), however, the bounding constants $C_U$ and $C_V$ in this case depend only on the background geometry $g_{ab}$.

\subsection{The Killing Laplacian and algebraic terms bound the first-order derivatives}
\label{Grad<Lap}

Combining our previous results (\ref{dLapTri}) and (\ref{HessKill}) bounds the $L^2$ norm of $\dLap v^a$ in terms of the $L^2$ norms of $\Lap v^a$, $\grad\!_a\, v^b$ and $v^a$ itself.  We now show that $\norm{\grad v}_{L^2(\M_\ind{}^\ind)}$ can also be bounded in terms of $\norm{\Lap v}_{L^2(\M^\ind)}$.  The idea is just to use the identity (\ref{KLWeitz}) and integration by parts to show that 
\begin{equation}
	\iprod{v}{\Lap v}_{L^2(\M^\ind)} 
		= \norm{\grad v}^2_{L^2(\M_\ind{}^\ind)} 
			+ \norm{\Div v}^2_{L^2(\M)} 
			+ \iprod{v}{\Ric v}_{L^2(\M^\ind)}.
\end{equation}
Throwing away the norm of the divergence and using the Schwarz inequality for the term involving the Ricci tensor gives the inequality we seek, 
\begin{equation}
	\norm{\grad v}^2_{L^2(\M_\ind{}^\ind)} 
		\le \iprod{v}{\Lap v}_{L^2(\M^\ind)} 
			+ \norm{v}_{L^2(\M^\ind)}\, \norm{\Ric v}_{L^2(\M^\ind)}.
\end{equation}
The expectation value on the right is non-negative because the Killing Laplacian is non-negative.  Furthermore, the term involving the Ricci tensor can be bounded via the argument that led to (\ref{Xbound}).  The result (\ref{GradLap}) follows.  Thus, the action of the perturbed Killing Laplacian on an arbitrary vector field $v^a$ can be bounded in terms of the action of the Killing Laplacian itself and terms algebraic in $v^a$.

\subsection{D\'enouement}
\label{dLap<Lap}

Nesting (\ref{HessKill}) and (\ref{GradLap}) inside (\ref{dLapTri}) gives the bound
\begin{align}\label{dLapLap}
	\norm[\big]{\dLap v}_{L^2(\M^\ind)} &\le 
		C_X\, \Bigl( 
			\norm{\Lap v}^2_{L^2(\M^\ind)}
			+ C_U\, \iprod{v}{\Lap v}_{L^2(\M^\ind, g)} 
			\\&\hspace{6em}\notag
			+ \bigl( C_V + C_U\, C_{\Ric} + C_{\Riem}^2 \bigr) \norm{v}^2_{L^2(\M^\ind)} \Bigr)^{1/2}
			\\&\hspace{2em}\notag
		+ C_Y\, \Bigl(
			\iprod{v}{\Lap v}_{L^2(\M^\ind, g)} 
			+ C_{\Ric}\, \norm{v}^2_{L^2(\M^\ind, g)} \Bigr)^{1/2}
		+ C_Z\, \norm{v}_{L^2(\M^\ind)}.
\end{align}
on the $L^2$ norm of $\dLap v^a$ for an arbitrary vector field $v^a$.  This result is of some interest in its own right, but our main objective now is to take $v^a = u_n^a$ to be an eigenfield of $\Lap$.   We then can combine the formal expression (\ref{dunform}) for the perturbed eigenfield $\dot u_n^a$ with the operator norm (\ref{pinorm}) of the pseudoinverse to yield (\ref{unBound}) with  
\begin{equation}\label{Cn}
	C_n := \frac{C_X \sqrt{\kappa_n^2 + C_U\, \kappa_n + C_V + C_U\, C_{\Ric} + C_{\Riem}^2}}{\gamma_n}
		+ \frac{C_Y \sqrt{\kappa_n + C_{\Ric}}}{\gamma_n}
		+ \frac{C_Z}{\gamma_n}
		+ \frac{C_A}{2}.
\end{equation}
The individual constants $C$ here are all of the form (\ref{CXdef}) for the tensor fields $X^{a \cdot bc}{}_d$, $Y^{a \cdot b}{}_c$ and $Z^a{}_b$ from (\ref{dkilllap}); $U_a{}^{b \cdot c}{}_d$ and $V^a{}_b$ from (\ref{bndHess}); the curvatures $R^a{}_b$ and $R_{a \cdot bc}{}^d$; and lastly $\dot\alpha^a{}_b$ from (\ref{alphadot}).  We have used a dot to separate indices, where necessary, to show how each of these tensor fields should be interpreted as a mapping between tensor products of tangent and cotangent spaces.  The grouping of indices is important for calculating the ``maximum maximum'' eigenvalue of each tensor field as on the first line (\ref{CXdef}), but not for merely calculating the trace on the second line.

We conclude this detailed discussion with three comments.

First, the constants $C_n$ from (\ref{Cn}) are homogeneous of degree one in the metric perturbation $\dot g_{ab}$ (and its derivatives).  Note that $X^{abc}{}_d$, $Y^{ab}{}_c$, and $Z^a{}_b$ from (\ref{dkilllap}), as well as $\dot\alpha^a{}_b$ from (\ref{alphadot}), scale proportionately under a (constant) rescaling of $\dot g_{ab}$.  The other ``$C$'' constants appearing in (\ref{Cn}) derive solely from the background geometry $g_{ab}$.

Second, the constants $C_n$ from (\ref{Cn}) vary with $n$, and increase like $\kappa_n / \gamma_n$ for large $n$.  One expects the eigenvalue gap $\gamma_n$ to increase with $n$, but not as rapidly as $\kappa_n$ itself.  Thus, for example, the fractional perturbations in the eigenvalues themselves, 
\begin{equation}\label{dlnkn}
	\frac{\dot\kappa_n}{\kappa_n}
		= \frac{\iprod[\nml]{u_n}{\dLap u_n}}{\kappa_n\, \norm{u_n}^2} 
		\le \frac{C_n}{\kappa_n},
\end{equation}
vanish for asymptotically large $n$.  Nonetheless, the $C_n$ themselves diverge in that limit.  Therefore, while each eigenfield $u_n^a$ varies continuously with $g_{ab}$ in the sense of (\ref{unBound}), the \textit{set} of eigenfields is not \textit{uniformly} continuous in $g_{ab}$.  This may raise some technical difficulties in certain applications, but our objective here is just to bound the variations only of the lowest eigenfield $u_0^a$.  Thus, what matters here is just the one constant $C_0$, and the subtlety regarding uniform continuity is irrelevant.

Third, we have shown that $u_0^a$ varies continuously with $g_{ab}$ in a loose sense, but have not stated clearly the topology on the space of metrics for which this continuity holds strictly.  Though this is a somewhat subtle issue, it certainly is clear that the relevant topology does not derive from a Hilbert-type ($L^2$) norm.  Rather, the constants $C_X$, $C_Y$, $C_Z$, and $C_A$ in (\ref{Cn}) arise by maximizing quantities over all of $\M$ as in (\ref{CXdef}).  The appropriate topology on the space of metrics should therefore be based on a supremum, or $L^\infty$-type, norm of $\dot g_{ab}$.  It should also have a Sobolev-type character, depending not only on $\dot g_{ab}$ but also on its first two derivatives.  The natural candidate is
\begin{equation}\label{sNorm}
	\norm{\dot g}^2_{W_2^\infty(\M_{(\ind\ind)}, g)} 
		:= \max_{p \in \M} \Bigl[ \dot g^{ab}\, \dot g_{ab} 
			+ \ell^2 \bigl( \grad\!_a \dot g_{bc} \bigr) \bigl( \grad^a \dot g^{bc} \bigr)
			+ \ell^4 \bigl( \grad\!_{(a} \grad\!_{b)} \dot g_{cd} \bigr) \bigl( \grad^{(a} \grad^{b)} \dot g^{cd} \bigr) 
			\Bigr], 
\end{equation}
where $\ell$ is a length scale that may depend on $g_{ab}$.  (For example, it might derive from the total volume of $\M$ in the background geometry.)  Conversely, we have established $L^2$-type bounds on the perturbations $\dot u_n^a$, but not $L^\infty$-type (\textit{i.e.}, supremum) bounds.  The latter are irrelevant, however, for calculating perturbations in quasi-local invariants like (\ref{BrownYork}) or (\ref{GenJ}).

\section{Conclusions}

This paper considers a family of smooth metrics $g_{ab}(\lambda)$ depending smoothly on a real parameter $\lambda$.  Each metric defines a Killing Laplacian $\Lap(\lambda)$, which admits a complete basis of eigenfields $u_n(\lambda)$ in the Hilbert space $L^2(\M^\ind, g)$ of vector fields on a compact manifold $\M$.  (The \textit{topology} of this Hilbert space, though of course not its inner product, is independent of the smooth metric $\M$ on the compact manifold $\M$.)  We have shown that each eigenfield $u_n^a(\lambda)$ varies continuously with $g_{ab}$ in the $L^2$ sense of (\ref{unBound}), where the size of a metric perturbation is measured in the $W_2^\infty$ sense of (\ref{sNorm}).  In particular, the lowest-order eigenfield $u_0^a$, which is a natural candidate to define an approximate Killing field on a manifold with no genuine symmetries, depends continuously on the metric.

The natural application of this result is to the problem of measuring the angular momentum of a black hole in general relativity.  The manifold $\M$ is a topological 2-sphere in this case, and the metric $g_{ab}$ on it is arbitrary.  The result is useful because in practical applications, say in numerical relativity, the actual horizon may be difficult or expensive to locate exactly.  Standard horizon-finding schemes \cite{Thornburg} help to locate the horizon approximately, and the continuous dependence of the approximate Killing eigenfield $u_0^a$ on the metric implies that the corresponding quasi-local angular momentum of the approximate horizon will approximate that of the actual horizon.  This conclusion is not unexpected, of course, but it is useful to establish a solid theoretical basis for the claim.  Similarly, the general continuity result we have established can be used to bound the growth of the angular momentum of a black hole under dynamic evolution.  This could be useful, for example, in the study of absorption processes involving dynamical horizons \cite{Ashtekar2004} at or near extremality.  Finally, in each of these applications we have proposed in (\ref{du0norm}) a novel, and purely local means of normalizing the approximate Killing field on a topological 2-sphere.

\subsection*{Acknowledgements}
The authors thank their colleagues at Florida Atlantic University for stimulating dicusssions of this work.  SW also gratefully acknowledges support under National Science Foundation Grant DGE:0638662 while part of this work was completed.

\section*{Bibliography}



\end{document}